\documentclass[a4paper]{revtex4}
\usepackage{graphicx}
\usepackage{fancyhdr}
\usepackage{amsmath}
\usepackage{color}

\begin{document}

\title{Inflation in a modified radiative seesaw model}


\author{Shoichi Kashiwase}
\affiliation{Institute for Theoretical Physics, Kanazawa University, Kanazawa 920-1192, JAPAN}

\begin{abstract}
A radiative seesaw model with an inert doublet dark matter is a promising candidate which could explain the existence of neutrino masses, dark matter and baryon number asymmetry of the Universe, simultaneously. 
In addition to these issues, inflation should also be explained since the recent CMB observations suggest the existence of the inflationary era at the early stage of the Universe. Thus, we extend it by a complex scalar field with a specific potential. This scaler could also be related to the neutrino mass generation at a TeV scale. We show that the inflation favored by the CMB observations could be realized even if inflaton takes sub-Plankian values during inflation.
\end{abstract}

\maketitle

\thispagestyle{fancy}


%%%%%%%%%%%%%%%%%%%%%%%%%%%%%%%%%%
\section{Introduction}
Recent discovery of a Higgs-like particle \cite{higgs} suggests that the framework of the standard model (SM) can describe Nature well up to the weak scale. On the other hand, we have experimental results which cannot be explained within it, that is, the existence of small neutrino masses 
\cite{t13}, the existence of dark matter (DM) \cite{uobs}, and baryon number asymmetry in the Universe \cite{baryon}. They require some extension of the SM. In our previous work, we show that a radiative seesaw model with an inert doublet \cite{Ma:2006km} could be a promising candidate which can explain these problems, simultaneously \cite{Kashiwase:2012xd}. 
The recent CMB observations suggest that the exponential expansion of the Universe occurs in the very early Universe. These results can constrain severely the allowed inflation models now \cite{Ade:2015lrj}. Some inflationary models require the trans-Planckian field value for inflaton to realize the sufficient e-foldings. In this case, the Planck-suppressed operators become dominant and spoil the flatness of the inflaton potential. Thus, we consider its modification to realize the results from Planck by introducing inflaton with the sub-Planckian field value. We also show that the inflaton could play a crucial role for the neutrino mass generation other than the inflation.
%%%%%%%%%%%%%%%%%%%%%%%%%%%%%%%%%%
\section{An extended model}
\begin{figure}[t]
\includegraphics[width=11cm,angle=0,clip]{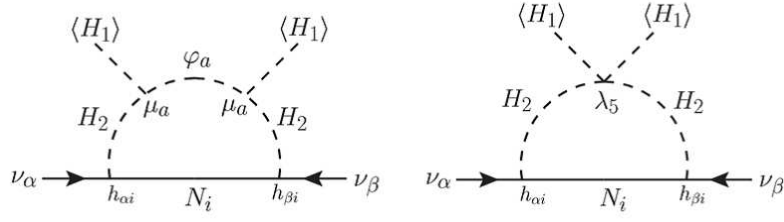}
\caption{One-loop diagrams which contribute to the neutrino mass generation. The left diagram generates neutrino masses in the present model. The dimensionful coupling $\mu_a$ is defined as 
$\mu_1=\frac{\mu}{\sqrt 2}$ and $\mu_2=\frac{i\mu}{\sqrt 2}$ by 
using $\mu$ in eq.~(\ref{lag}). Right one is the diagram in the original 
model \cite{Ma:2006km}. \label{ndgm}}
\end{figure}

The extended radiative seesaw model with a $Z_2$ odd complex scalar singlet $S$ is defined by the following $Z_2$ invariant terms \cite{Budhi:2014gxa}:
\begin{align}
\nonumber -{\cal L}_{O}&=-h_{\alpha i}\bar N_iH_2^\dagger L_\alpha-h_{\alpha i}^*\bar L_\alpha H_2 N_i+\frac{1}{2}m_{N_i}\bar N_iN_i^c+\frac{1}{2}m_{N_i}^*\bar N_i^cN_i,\\
\nonumber&+m_1^2|H_1|^2+m_2^2|H_2|^2+\lambda_1|H_1|^4+\lambda_2|H_2|^4+\lambda_3|H_1|^2|H_2|^2+\lambda_4|H_1^{\dagger}H_2|^2+\frac{\lambda_5}{2} \left[(H_1^{\dagger}H_2)^2+h.c.\right],\\
\nonumber-{\cal L}_{S}&=\tilde m_S^2S^\dagger S+\frac{1}{2}m_S^2S^2+\frac{1}{2}m_S^2{S^\dagger}^2-\mu SH_2^\dagger H_1-\mu^*S^\dagger H_1^\dagger H_2\\
&+\kappa_1(S^\dagger S)^2+\kappa_2(S^\dagger S)(H_1^\dagger H_1)+\kappa_3 (S^\dagger S)(H_2^\dagger H_2), \label{lag}
\end{align}
where $L_\alpha$ is a left-handed lepton doublet and $H_2$ is an inert doublet scalar. Since $H_2$ and right-handed neutrinos $N_i$ are assigned odd parity of $Z_2$ symmetry and all the SM contents including the ordinary Higgs doublet scalar $H_1$ are assigned even parity, neutrino Dirac mass terms are forbidden at tree level. Neutrino masses are generated through a one-loop diagram as shown in Fig.~\ref{ndgm}. 
In this diagram, $\varphi_a$ represents component fields of $S$ which are defined as $S=\frac{1}{\sqrt{2}}(\varphi_1+\varphi_2)$. Their masses are found to be $\bar m_1^2=\tilde m_S^2+m_S^2$ and $\bar m_2^2=\tilde m_S^2-m_S^2$. Since $Z_2$ is considered as an exact symmetry, $\tilde m_S^2>m_S^2$ should be satisfied. If we assume the condition $\tilde m_S\gg m_S,m_2,m_{N_i}$ is satisfied, the neutrino mass induced through the diagram in Fig.~\ref{ndgm} can be estimated as 
\begin{align}
{\cal M}^\nu _{\alpha\beta}=\sum_{i=1}^3h_{\alpha i}h_{\beta i}\frac{\langle H_1\rangle^2}{8\pi ^2}\frac{m_S^2\mu^2}{\tilde m_S^4}\frac{m_{N_i}}{m_{H_2}^2-m_{N_i}^2}\left(1+\frac{m_{N_i}^2}{m_{H_2}^2-m_{N_i}^2}\ln\frac{m_{N_i}^2}{m_{H_2}^2}\right),
\end{align}
where $m_{H_2}^2=m_2^2+(\lambda_3+\lambda_4)\langle H_1\rangle^2$. It is equivalent to the neutrino mass formula in the original model if $\frac{m_S^2\mu^2}{\tilde m_S^4}$ is identified with the coupling constant $\lambda_5$ for a $(H_2^\dagger H_1)^2$ term.

This correspondence might be found in an effective theory obtained at energy regions smaller than $\tilde m_S$ by integrating out $S$. The origin of small $\lambda_5$ which is the key nature to explain the 
smallness of the neutrino masses is now translated to the hierarchy 
problem between $\mu$, $m_S$ and $\tilde m_S$ in this extension.
If we leave the origin of this hierarchy to a complete theory 
at high energy regions, all the neutrino masses, the DM abundance 
and the baryon number asymmetry could be also explained in this extended 
model at TeV regions just as in the same way discussed in the 
previous articles \cite{Kashiwase:2012xd}. 
Following the results obtained in these studies, the value of 
$\frac{m_S^2\mu^2}{\tilde m_S^2}$ could be constrained by the simultaneous 
explanation of these.
\section{Inflation due to the complex scalar $S$}
We consider an inflation scenario which could work even for sub-Plankian values of $S$, following proposal in \cite{McDonald:2014oza}. It is possible as long as the existence of specific nonrenormalizable terms is assumed in the potential for $S$.
As such a potential, we suppose that the complex scalar $S$ has $Z_2$ 
invariant additional terms such as 
\begin{align}
V&= c_1\frac{(S^\dagger S)^n}{M_{\rm pl}^{2n-4}}
\left[1+ c_2\left\{ \left(\frac{S}{M_{\rm pl}}\right)^{2m} 
\exp\left(i\frac{S^\dagger S}{\Lambda^2}\right)
+ \left(\frac{S^\dagger}{M_{\rm pl}}\right)^{2m}
\exp\left(-i\frac{S^\dagger S}{\Lambda^2}\right)
\right\} \right], \nonumber \\
&=c_1\frac{\varphi^{2n}}{2^nM_{\rm pl}^{2n-4}}\left[
1+ 2c_2\left(\frac{\varphi}{\sqrt 2 M_{\rm pl}}\right)^{2m}
\cos\left(\frac{\varphi^2}{2\Lambda^2}+2m\theta\right)\right],  
\label{model3}
\end{align}
where both $n$ and $m$ are positive integers 
and $M_{\rm pl}$ is the reduced Planck mass. Here we assume the condition $\Lambda\ll\varphi\ll M_{\rm pl}$. We use the polar coordinate expression $S=\frac{\varphi}{\sqrt 2}
 e^{i\theta}$ in the second equality of eq.~(\ref{model3}). 
\begin{figure}[b]
\begin{center}
\includegraphics[width=6cm,angle=0,clip]{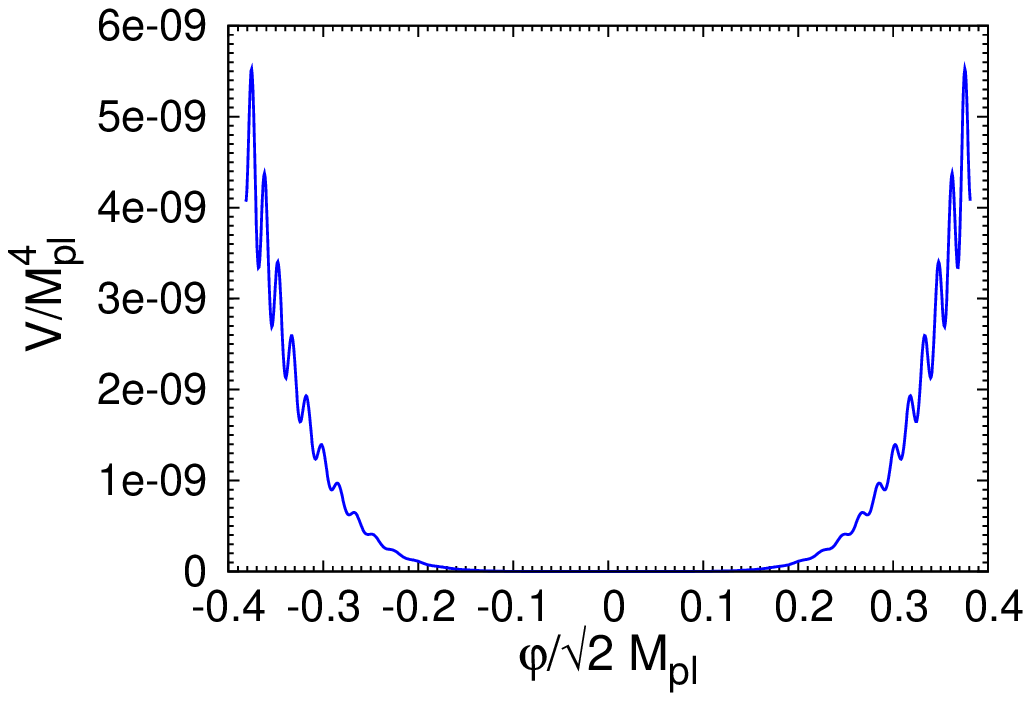}
\hspace{1mm}
\includegraphics[width=6cm,clip]{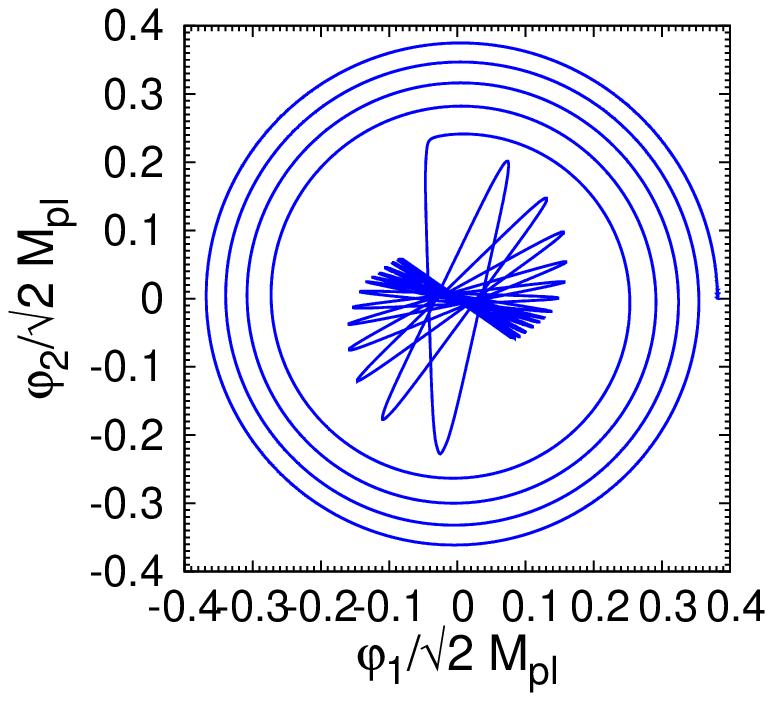}
\caption{The left panel shows the potential $V$ defined by
$n=3$ and $m=1$. Other Parameters in $V$ are fixed as $c_1=1.65\times 10^{-6}$, 
$c_2=0.7$ and $\Lambda/M_{\rm pl}=0.04$. The right panel depicts 
the time evolution of the inflaton in the 
$(\frac{\varphi_1}{\sqrt 2M_{\rm pl}},\frac{\varphi_2}{\sqrt 2M_{\rm pl}})$ plane 
for the potential $V$ shown in the left panel. $\varphi$ is related to
$\varphi_{1,2}$ by $\varphi^2=\varphi_1^2+\varphi_2^2$. }
\label{tra}
\end{center}
\end{figure}
In the left panel of Fig.~\ref{tra}, we show a typical shape of 
the potential as a function of $\varphi$ for a fixed $\theta$.
We assume that inflaton moves along this local minimum. In the right panel of Fig.~2, we show an example for the evolution of the inflaton in the 
$(\frac{\varphi_1}{\sqrt 2M_{\rm pl}}, \frac{\varphi_2}{\sqrt 2M_{\rm pl}})$ 
plane. In this calculation, we assume that $\varphi_{1,2}$ initially stay at the local minimum. The figure shows that the inflaton evolves along an aperiodic circle.
During this evolution, the value of inflaton changes by an amount larger
than the Planck scale for the small change of $\varphi$
in the sub-Planckian range. From this figure, we find that the single inflaton scenario could be realized in this model as long as we assume that the conditions mentioned above are satisfied and also the fields $\varphi_{1,2}$ start to evolve 
from a local minimum.

\begin{figure}[b]
\includegraphics[width=11.3cm,angle=0,clip]{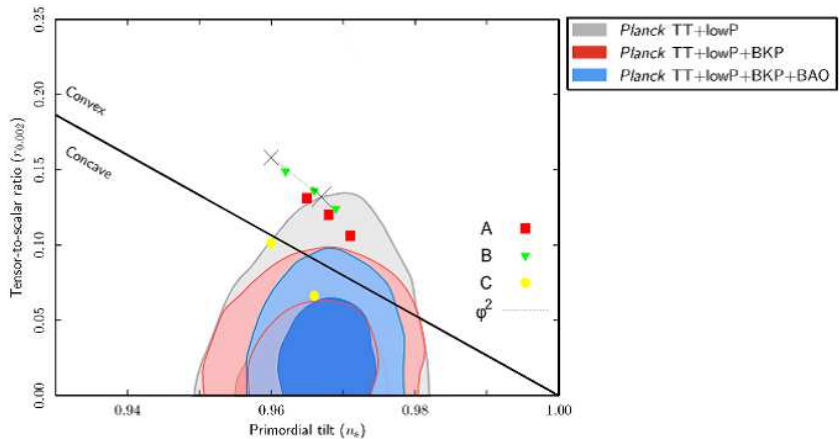}
\caption{Predicted values of $(n_s, r)$ for several 
parameter sets $(c_2,\frac{\Lambda}{M_{\rm pl}})$ given in Table 1. 
A dotted line represents the prediction by the quadratic chaotic inflation 
model and the crosses correspond to the points for $N_\ast=50$ and 60. 
Contours given as Fig.~54 in Planck Collaboration XX \cite{Ade:2015lrj} are 
used here.}\label{cmb}
\end{figure}

In order to see the feature of the inflation in this model,
we calculate the quantities which characterize the inflation, that is, 
the e-foldings $N$, the spectral index $n_s$ and the tensor-to-scalar ratio 
$r$ (See more details in \cite{Budhi:2014gxa}). In Table~1, we show typical examples which are calculated numerically for different values for the model parameters $c_1$, $c_2$ and 
$\Lambda$. These examples suggest that sufficiently large e-foldings 
such as $N_\ast=50$ - 60 could be realized as long as $\Lambda \ll \varphi_\ast$ is satisfied even for the sub-Planckian inflaton value 
$\varphi_\ast< M_{\rm pl}$. The predicted values of $n_s$ and $r$ are also listed in each case. In Fig. \ref{cmb}, we plot the predicted points in the $(n_s, r)$ plane for $N_\ast=50$ - 60 in the cases A, B and C given in Table 1. As we can see from this figure, both cases A and B which were favored by BICEP2 \cite{Ade:2014xna} have been excluded by the recent Planck data \cite{Ade:2015lrj}. On the other hand, the case C is in the region of the 95\% CL due to the Planck results. We will examine the viable region in this model more extensively in \cite{pre}.   

\begin{table}[t]
\begin{center}
\caption{The spectral index and the 
tensor-to-scalar ratio for typical examples of three parameters in the potential (3) defined by $n=3$ and $m=1$. 
These model parameters are fixed to realize the observed value 
for the scalar perturbation amplitude $\Delta_{\cal R}^2$ at $k_*=0.002$ Mpc$^{-1}$.}
\begin{tabular}{|c|c|c|c|c||c|c|c|}
\hline
&$c_1$&$c_2$&$\frac{\Lambda}{M_{\rm pl}}$&$\frac{\varphi_*}{\sqrt2M_{\rm pl}}$&$N_*$&$n_s$&$r$\\ \hline
A&$1.66\times10^{-6}$&0.7&0.04&0.378&59.0&0.971&0.107\\ 
  &$2.04\times10^{-6}$&0.7&0.04&0.371&54.2&0.968&0.119\\ 
  &$2.42\times10^{-6}$&0.7&0.04&0.366&49.1&0.965&0.131\\ 
\hline
B&$0.257$&6.0&0.002&0.0512&60.4&0.969&0.124\\ 
  &$0.305$&6.0&0.002&0.0505&55.0&0.966&0.136\\ 
  &$0.364$&6.0&0.002&0.0498&50.0&0.962&0.149\\ 
\hline
C&$0.82\times10^{-6}$&1.4&0.05&0.425&66.6&0.966&0.066\\ 
  &$1.52\times10^{-6}$&1.4&0.05&0.406&48.3&0.960&0.101\\ 
\hline
\end{tabular}
\label{example_table}
\end{center}
\end{table}

\section{Reheating after the end of inflation}
In this model, the Universe could be reheated up through inflaton decay after the end of inflation. Since $\tilde m_S\gg m_S$ is assumed to be satisfied here, the reheating temperature realized through this process could be estimated as
\begin{equation}
T_R\simeq 1.6\times 10^8\left(\frac{|\lambda_5|}{10^{-6}}\right)^{1/2}
\left(\frac{\tilde m_S}{m_S}\right)
\left(\frac{\tilde m_S}{10^6~{\rm GeV}}\right)^{1/2}~{\rm GeV}. 
\label{reheat}
\end{equation}
In this estimation, we take account of the constraint from the neutrino mass generation as discussed in the previous part. Here we also note that $|\lambda_5|$ should be larger 
than $O(10^{-6})$, which is imposed by the present bound of DM direct search 
since we suppose that the lightest neutral component of $H_2$ is DM
and its mass is $\sim 1$~TeV \cite{Kashiwase:2012xd}.
We find that the reheating temperature could take values in a wide range
such as $10^5{\rm GeV}~{^<_\sim}~T_R~{^<_\sim}~10^{15}~{\rm GeV}$ depending on 
a value of $\tilde m_S$.
This temperature is high enough to produce thermal right-handed neutrinos
in the present model since the masses of right-handed neutrinos are assumed 
to be of $O(1)$ TeV. 
If the right-handed neutrino masses are sufficiently degenerate, 
the baryon number asymmetry could be generated through
the resonant leptogenesis as discussed in \cite{Kashiwase:2012xd}. 
Right-handed neutrinos need not to be light but they could have large 
mass such as $O(10^9)$ GeV in a consistent way with this 
neutrino mass model \cite{Kashiwase:2012xd}. 
Even in that case, eq.~(\ref{reheat}) shows that the reheating 
temperature could be high enough for leptogenesis to work well without 
the resonant effect.
\section{conclusion}
We have considered an extension of the radiative seesaw 
model with a complex singlet scalar to realize the inflation of 
the Universe keeping favorable features of the original model,
that is, the simultaneous explanation of the small neutrino masses,
the DM abundance and the baryon number asymmetry in the Universe.
This singlet scalar plays a crucial role not only 
for inflation but also for the small neutrino mass generation.
In this scenario, the inflaton trajectory follows an aperiodic circle during the 
inflation. This feature makes it possible that sub-Planckian values of the relevant field induce trans-Planckian change of the inflaton value which is needed for the sufficient e-foldings. The model could be free from the serious 
problem caused by trans-Planckian field values. Both the spectral index $n_s$ and the tensor-to-scalar ratio $r$ could have values which are favorable from the recent CMB observations. 
The roughly estimated reheating temperature could be high enough for leptogenesis.
\begin{acknowledgments}
This work is collaboration with Romy H. S. Budhi and Daijiro Suematsu. The author would like to thank them for their support. S.~K. is supported by Grant-in-Aid for JSPS fellows (26$\cdot$5862).
\end{acknowledgments}

\bigskip % extra skip inserted

\end{document}